\documentclass[a4paper]{article}
\usepackage{indentfirst}
\usepackage{hyperref}
\usepackage{multirow}
\usepackage{latexsym,bm,amsmath,amssymb}
\usepackage{graphicx}
\usepackage{epsfig}
\usepackage{subfigure}
\usepackage{cite}
\usepackage{color}
\usepackage[top=1in, bottom=1in, left=1.25in, right=1.25in]{geometry}
\usepackage{slashed}
\usepackage{fancyhdr}
\usepackage{slashed}
\usepackage{makecell,rotating,multirow,diagbox}
\usepackage{tikz}
\usepackage{float}
\usepackage{appendix}
\usepackage{setspace}
\usepackage{geometry}
\usepackage{graphics}
\usepackage{multirow}
\usepackage{booktabs}
\usetikzlibrary{shapes.geometric, arrows}
\usetikzlibrary{trees}
\usetikzlibrary{decorations.pathmorphing}
\usetikzlibrary{decorations.markings}
\tikzset{
        photon/.style={decorate, decoration={snake}, draw=red},
        nucleon/.style={draw=black, postaction={decorate},
           decoration={markings,mark=at position .55 with{\arrow[draw=black]{>}}}},
        pion/.style={draw=blue, postaction={decorate},
        decoration={markings,mark=at position .55 with{\arrow[draw=blue]{}}}},
        sigma/.style={draw=black, postaction={decorate},
        decoration={markings,mark=at position .55 with{\arrow[draw=black]{}}}},
        link/.style    = { draw=black, double = white, line width = 1.8pt, double distance = 0.8pt,
        postaction={decorate},decoration={markings,mark=at position .55 with{\arrow[draw=black]{>}}}},
    }

\usepackage{lipsum} 
\newlength{\halfpagewidth}
\newcommand{\Rone}{\uppercase\expandafter{\romannumeral1}}
\newcommand{\Rtwo}{\uppercase\expandafter{\romannumeral2}}
\newcommand{\Rthree}{\uppercase\expandafter{\romannumeral3}}
\newcommand{\Rfour}{\uppercase\expandafter{\romannumeral4}}

\newcommand{\jpsi}{J/\psi}

\begin{document}
\title{A discussion on the  anomalous threshold enhancement of $J/\psi$ -- $\psi(3770)$ couplings and  $X(6900)$ peak}
\maketitle

\begin{center}
{\sc
Ye Lu$^{\dagger\,}$,\,
Chang Chen$^{\dagger\dagger\,}$,\,
Guang-you Qin$^{\dagger\,}$,\,
Han-Qing~Zheng$^{\heartsuit\,}$}
\\
\vspace{0.5cm}
\noindent{\small{$^{\dagger}$ \it  Institute of Particle Physics and Key Laboratory of Quark and Lepton Physics (MOE), Central China Normal University, Wuhan, Hubei 430079, China}}\\
\noindent{\small{$^{\dagger\dagger}$ \it  Department of Physics,
Peking University, Beijing 100871, China}}\\
\noindent{\small{$^\heartsuit$ \it College of Physics, Sichuan University, Chengdu, Sichuan 610065,  China}}\\
\end{center}
\begin{abstract}
The attractive interaction between $J/\psi$ and $\psi(3770)$ has to be strong enough to form the molecular resonant structure, i.e., $X(6900)$. We argue that 
since $\psi(3770)$ decays predominantly into a $D\bar D$ pair, the interactions between  $J/\psi$ and $\psi(3770)$ may be significantly enhanced due to the three point $D\bar D$ loop diagram. The enhancement comes from the anomalous threshold located at $t=-1.288$GeV$^2$, which effect propagates into the $s$-channel partial wave amplitude in the vicinity of  $\sqrt{s}\simeq 6.9$GeV. This effect  may be helpful in forming the $X(6900)$ peak.
\end{abstract}


%

The  $X(6900)$ peak observed by the  LHCb Collaboration  in the di-$\jpsi$ invariant mass spectrum~\cite{6900}\cite{CMS},  and later in the  $J\psi\psi(3686)$ invariant mass spectrum~\cite{ATLAS:2022hhx}.  has stimulated many discussions in theory aspects (see for example Ref.~\cite{Lu:2023ccs} for an incomplete list of references). On the other side,  since $X(6900)$  is close to the threshold of $\jpsi \psi(3770)$, $\jpsi \psi_2(3823)$, $\jpsi$$\psi_3(3842)$, and $\chi_{c0} \chi_{c1}$ (and  $X(7200)$ is close to the threshold of  $\jpsi \psi(4160)$ and $\chi_{c0}$$\chi_{c1}(3872)$), 
inspired by this, it is studied in Ref.~\cite{Cao:2020gul} the properties of $X(6900)$ and $X(7200)$, by assuming the $X(6900)$ coupling to $\jpsi\jpsi$, $\jpsi\psi(3770)$, $\jpsi \psi_2(3823)$, $\jpsi \psi_3(3842)$ and $\chi_{c0}\chi_{c1}$ channels, and  $X(7200)$ to  $\jpsi\jpsi$, $\jpsi\psi(4160)$ and $\chi_{c0}\chi_{c1}(3872)$ channels.
For the $S$-wave $\jpsi \jpsi$ coupling, the pole counting rule (PCR)~\cite{pole}, which has been applied to the studies of ``$XYZ$" physics in Refs.~\cite{Zhang:2009bv,Dai:2012pb,X3900,Cao:2019wwt}
is employed to analyze the nature of the two structures. It is found that the di-$J/\psi$ data alone is not enough to judge the intrinsic properties of the two states. It is also pointed out that the $X(6900)$ is unlikely a molecule of $J/\psi\,\psi(3686)$~\cite{Cao:2020gul} -- a conclusion drawn before the discovery of Ref.~\cite{ATLAS:2022hhx}. More recently, Refs.~\cite{Liang:2021fzr, Zhou:2022xpd,Lu:2023ccs} made an analysis on $X(6900)$ using a combined analysis on di-$J/\psi$ and $J/\psi\psi(3686)$ data and concluded that $X(6900)$ cannot be a $J/\psi\psi(3686)$ molecule. 

Nevertheless, as already stressed in Ref.~\cite{Lu:2023ccs}, even though  $X(6900)$ is very unlikely a molecule of  $J/\psi\psi(3686)$, it does not mean that it has to be an ``elementary state" (i.e., a compact $\bar c\bar ccc$ tetraquark state).  It is pointed out  that, it is possible that  $X(6900)$ be a molecular state composed of other particles, such as $\jpsi\psi(3770)$,  which form thresholds closer to $X(6900)$, if the channel coupling is sufficiently large.

\begin{figure}
    \centering
    \includegraphics[width=0.7\linewidth]{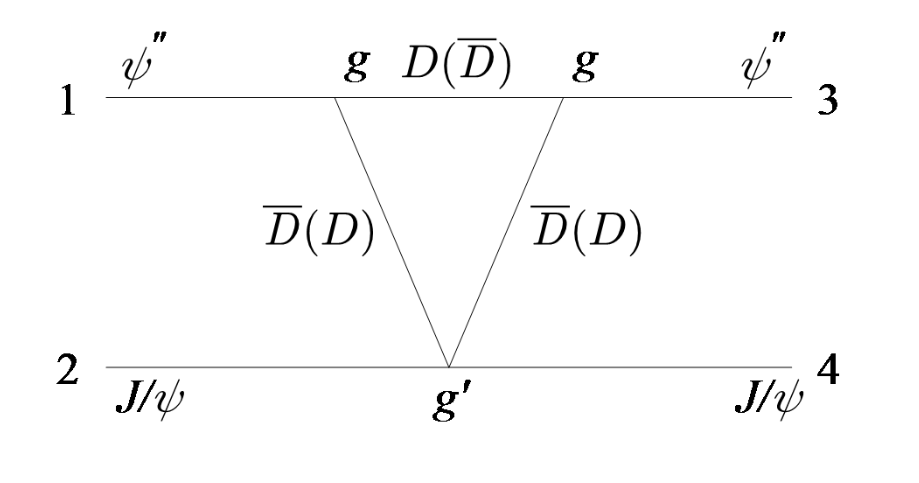}
   \caption{The $D\bar D$ triangle diagram in $J/\psi \psi''$ scattering process.}
    \label{fig:1}
\end{figure}

This note will discuss a possible mechanism for the enhancement of the $\jpsi\psi(3770)$ channel coupling.  The $D\bar D$ (or $D^*\bar D^*$) component inside $\psi(3770)$ may play an important role, so far ignored in the literature, in explaining the  $X(6900)$  resonant peak, through the anomalous threshold emerged from the triangle diagram generated by $D$ ($D^*$) loop, as depicted in Fig.~\ref{fig:1}. 

\begin{figure}[H]
    \centering
    \subfigure[]{
    \includegraphics[scale = 0.45]{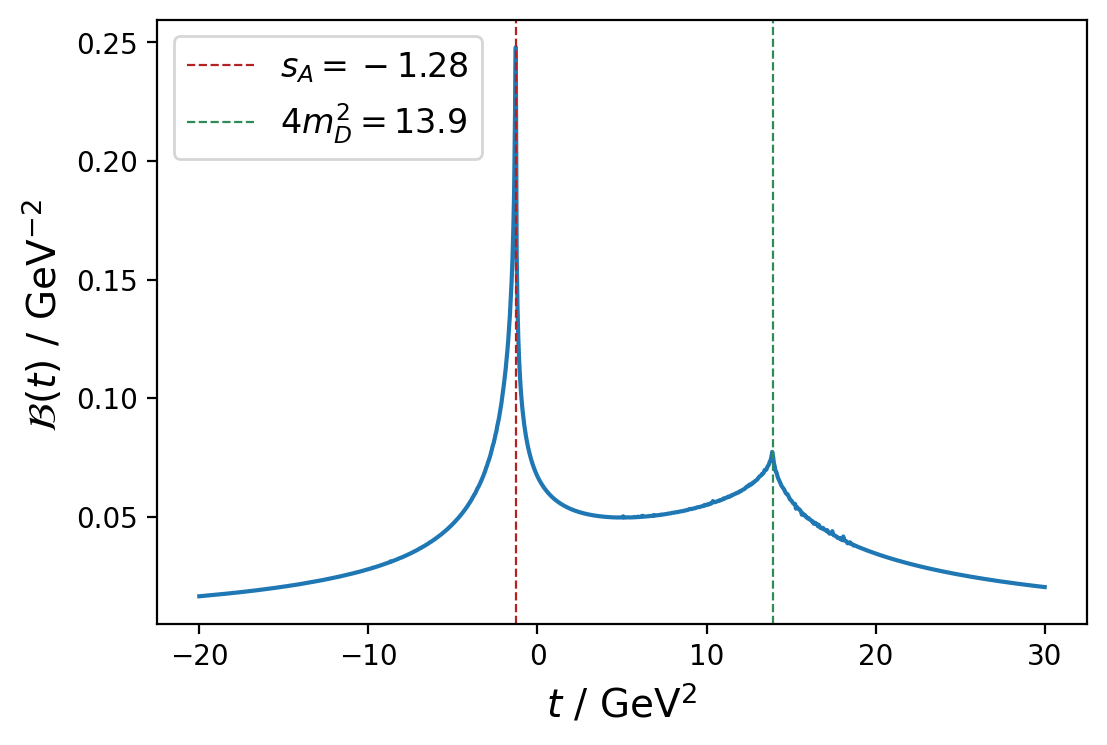}
    }
    \subfigure[]{
    \includegraphics[scale = 0.51]{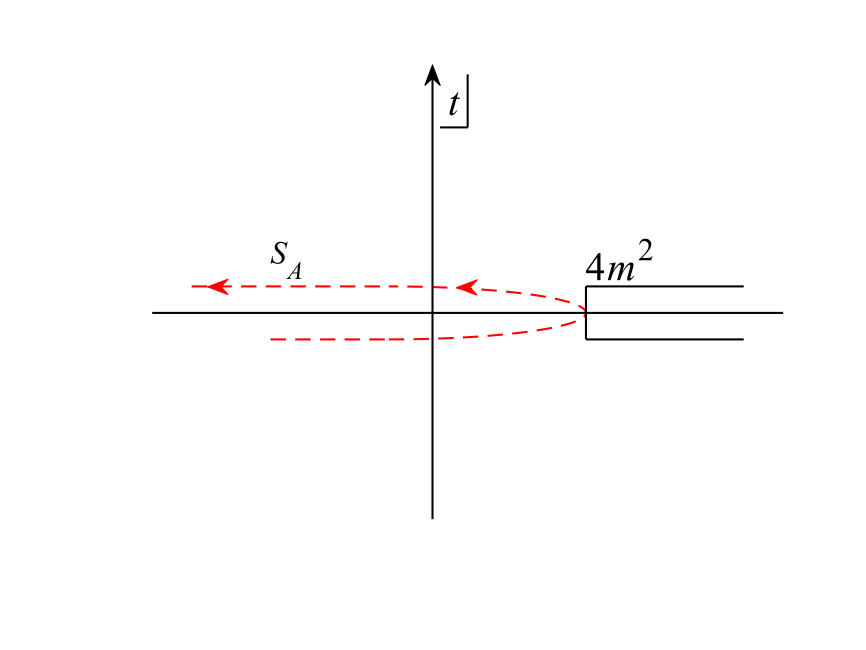}
    }
    \caption{Left: triangle diagram contribution (the y axis label is arbitrary). Right: the trajectory of the anomalous threshold with respect to the variation of $M^2$.   }
    \label{fig:2}
\end{figure}

Noticing that $\psi(3770)$ or $\psi''$ couples dominantly to $D\bar D$, we start from the Feynman diagram as depicted in Fig.~\ref{fig:1} by assuming it contributes $J/\psi\psi''$ elastic scatterings near  $J/\psi \psi''$ threshold. \footnote{It could be wondered that the $X(6900)$ be a $\bar DD J/\psi$ three body halo state, see Refs.~\cite{Braaten, Hammer}.} 
Assuming an interaction lagrangian\footnote{We neglect all the complexities such as form factors, hence our calculations are only qualitative or at best semi-quantitative.}
\begin{equation}
\begin{aligned}
     \mathcal{L} =& -ig(D^0\partial_\mu\bar{D}^0 - \bar{D}^0\partial_\mu D^0) \psi''^\mu -ig (D^+\partial_\mu D^- - D^-\partial_\mu D^+) \psi''^\mu\\
    &+ g' D^0\bar{D}^0 J/\psi^\mu J/\psi_\mu+ g' D^+D^- J/\psi^\mu J/\psi_\mu\ ,
\end{aligned}
\end{equation}
after performing the momentum integration the amplitude as depicted by Fig.~\ref{fig:1} is
\begin{equation}
\begin{aligned}
    i\mathcal{M}& = (-16 g^2 g') ~ (\epsilon_2 \cdot \epsilon_4) ~\epsilon_1^\mu \epsilon_3^\nu \int d^Dk \frac{k_\mu k_\nu}{((k-p_1)^2-m_D^2)((k-p_3)^2-m_D^2)(k^2-m_D^2)}\\
    & \equiv (-16 g^2 g') ~ (\epsilon_2 \cdot \epsilon_4) ~\epsilon_1^\mu \epsilon_3^\nu ~\mathcal{A}_{\mu\nu}
\end{aligned}
\end{equation}
where
\begin{equation}\label{A}
\begin{aligned}
    \mathcal{A}^{\mu\nu}& =\frac{-i}{16\pi^2}\int_0^1 dx\int_0^{1-x}dy ~ \left\{\frac{p_3^\mu p_1^\nu ~ xy}{\Delta} - \frac{g^{\mu\nu}}{4} \Gamma(\epsilon) \frac{1}{\Delta^\epsilon} \right\}\\
    & \equiv \frac{-i}{16\pi^2}   \left\{  p_3^\mu p_1^\nu~ \mathcal{B} +    \int_0^1 dx\int_0^{1-x}dy \left(- \frac{g^{\mu\nu}}{4} \Gamma(\epsilon) \frac{1}{\Delta^\epsilon} \right)\right\}\ ,
\end{aligned}
\end{equation}
and
\begin{equation}\label{master}
\begin{aligned}
    \mathcal{B}(t) 
              = &\int_0^1 dx \frac{x}{2M^2} \left( 2(M^2(1-2x)+tx)\frac{\mathrm{ArcTan}(\frac{M^2-tx}{\Lambda(t,x)})-\mathrm{ArcTan}(\frac{M^2(2x-1)-tx}{\Lambda(t,x)})}{\Lambda(t,x)} \right.\\
              &+\left. \ln \frac{m^2+t(x-1)x}{m^2+M^2(x-1)x}\right)\ .
\end{aligned}
\end{equation}
In above $\Lambda(t, x)= \sqrt{
 4 m^2 M^2 - M^4 + 4 M^2 t x^2 - 2 M^2 t x - t^2 x^2}$,  { $\Delta = M^2(x^2+y^2)+(2M^2-t)xy-M^2(x+y)+m^2$,
$\Gamma(\epsilon) \frac{1}{\Delta^\epsilon} = \frac{1}{\epsilon} - \ln \Delta - \gamma + \ln 4\pi + \mathcal{O}(\epsilon)$.}
On the right hand side of Eq.~(\ref{A}), only the $\mathcal{B}$ term will be considered since the rest will be absorbed by the contact interactions to be introduced latter. $M$ is the mass of $\psi(3770)$, $m$ is the mass of $D(\bar D)$.
The parameter $g$ is the coupling strength of $\psi''$ $D\bar D$ three point vertex, 
 $g’$ the coupling strength of $J/\psi J/\psi D\bar D$ four point vertex; {$t=(p_2-p_4)^2$}. 
The parameter $g$ can be determined by the process $\psi''\to D^0\bar{D}^0$ decay, 
\begin{equation}
\begin{aligned}
    &i \mathcal{M}_{\psi'' DD} = i \,g \,\epsilon(\psi'') \cdot [p(D^0)-p(\bar{D}^0)]\ ,\\
    &\Gamma = \frac{1}{8\pi}\,|i \mathcal{M}_{\psi'' DD}|^2\,\frac{q(DD)}{M_{\psi''}^2} = \frac{1}{6 \pi} \, g^2 \, \frac{q(DD)^3}{M_{\psi''}^2}\ ,
\end{aligned}
\end{equation}
where $q(DD)$ is the norm of three-dimensional momentum of $D^0$ or $\bar{D}^0$ in final state. The PDG value $\Gamma_{\psi''\to D^0\bar{D}^0} \sim 27.2\times 52\% \times10^{-3}$~\cite{PDG} determines $g \sim 12$. \footnote{In Ref.~\cite{Coito:2017ppc}, g is estimated to be larger ($\sim 30$).} Parameter $g'$ is unknown and left as a free parameter.
The amplitude Eq.~(\ref{master}) contains rather complicated singularity structure, especially the well-known anomalous threshold, which in history was discovered by Mandelstam who use it to explain the looseness  of deuteron wave function~\cite{Mandelstam}. The anomalous threshold locates at
\begin{equation}
    s_A = 4m^2-\frac{(M^2-2m^2 )^2 }{m^2}\ .
\end{equation}
Putting the mass of $\psi''$ and $D^0$ mesons, one gets $s_A=-1.28$GeV$^2$ (for $D^+$ loop,  -0.98GeV$^2$ ). Numerically function $\cal{B}$ is plotted in Fig.~\ref{fig:2}a, where one clearly sees the anomalous threshold, beside the normal one at $t=4m^2$.
Notice that if $M^2$ were smaller than $2m^2$, the anomalous branch point locates below physical threshold, but on the second sheet. It touches the physical threshold $4m^2$ and turns up to the physical sheet if the value of $M^2$ increased to $2m^2$. When further increase $M^2$, the anomalous threshold moves towards left on the real axis  and pass the origin when $M^2=4m^2$, and finally reach the physical value, i.e.,   -1.28GeV$^2$. The situation is depicted in Fig.~\ref{fig:2}b. Notice that $s_A$ here is negative, contrary to what happens in  deuteron, because the latter is a bound state with a normalizable wave function,  whereas $\psi''$ is an unstable resonance.   

\begin{figure}
    \centering
    \includegraphics[width=0.5\linewidth]{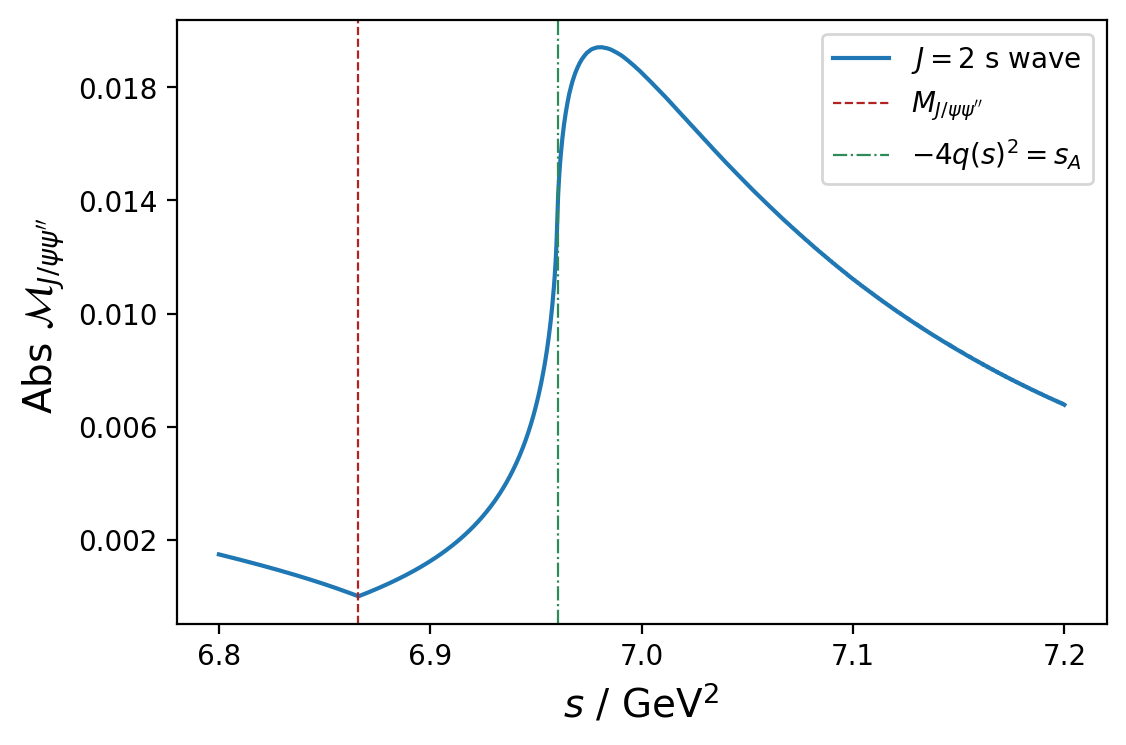}
    \caption{The enhancement of the $J/\psi\psi''$ scattering amplitude from the triangle diagram.}
    \label{fig:3}
\end{figure}

To proceed, one needs further to make the partial wave projection of $\mathcal{M}$ and gets
\begin{equation}\label{eq1}
    T^J_{\mu_1\mu_2\mu_3\mu_4}(s)=\frac{1}{32\pi}\frac{1}{2q^2(s)}\int^0_{-4q^2(s)}\,dt\, d^J_{\mu\mu'}\left(1+\frac{t}{2q^2(s)}\right)\mathcal{M}_{\mu_1\mu_2\mu_3\mu_4}(t)\ ,
\end{equation}
where the channel momentum square reads: $q^2(s)=(s - (M - M_J)^2) (s - (M + M_J)^2)/4s$ with $M_J$ the mass of $J/\psi$, $\mu_i$ denotes the corresponding
helicity configuration and $\mu = \mu_1-\mu_2$, $\mu' = \mu_3-\mu_4$. The key observation is  that the integral interval in Eq.~(\ref{eq1}) will cover  $s_A$ if $\sqrt{s}>6.96$GeV (for $D^+$ loop,  $\sqrt{s}>6.94$GeV). In other words, the partial wave amplitude will be  enhanced in the vicinity of $X(6900)$ peak by the 
anomalous threshold enhancement in the $t$ channel, see Fig.~\ref{fig:3}  for illustration.

\begin{figure}
    \centering
    \includegraphics[width=1.\linewidth]{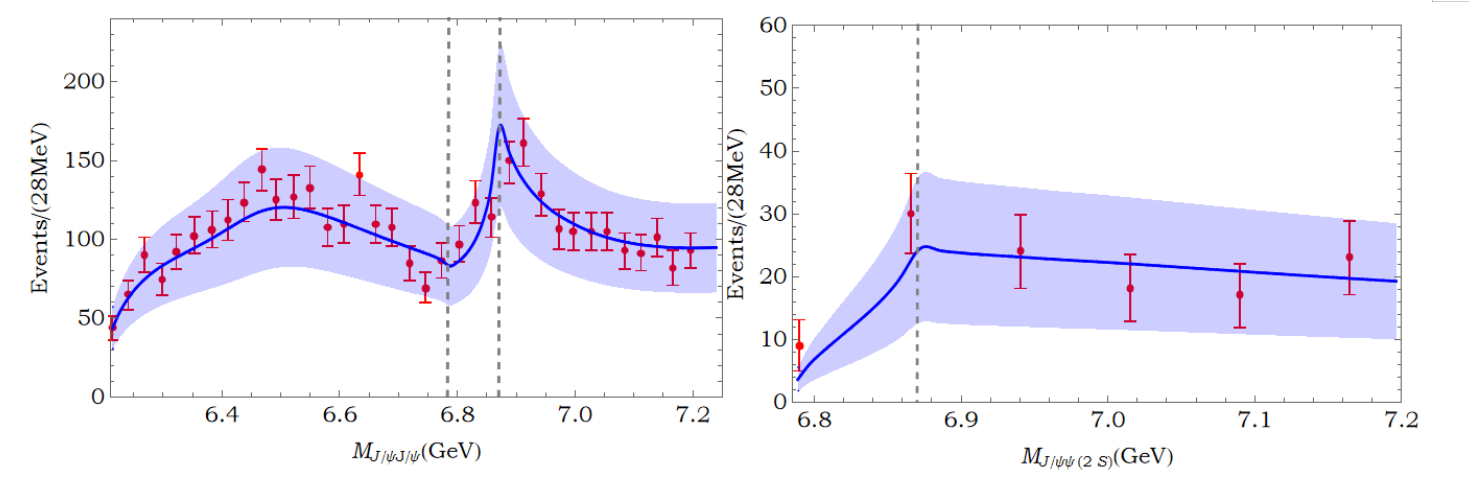}
    \caption{The fit results from table~1.}
    \label{fit}
\end{figure}

Based on the above observation it is  suggested that the $X(6900)$ peak may at least partly be explained by the anomalous threshold generated by the triangle diagram as depicted in Fig.~\ref{fig:1}. 
Furthermore, to get the $L = 0$ (s wave) amplitudes, we need the relation between s wave and helicity amplitudes (see Refs.~\cite{Liang:2021fzr, Zhou:2022xpd} for more discussions):
\begin{equation}
T_{L=0}^{0,i j}(s)=\frac{1}{3}\left[2 T_{++++}^{0, i j}(s)+2 T_{++--}^{0, i j}(s)-2 T_{++00}^{0, i j}(s)-2 T_{00++}^{0, i j}(s)+T_{0000}^{0, i j}(s)\right],
\end{equation}
\begin{equation}
\begin{aligned}
T_{L=0}^{2, i j}(s)&=  \frac{1}{15}[T_{++++}^{2, i j}(s)+T_{++--}^{2, i j}(s)]\\
&+\frac{\sqrt{6}}{15}[T_{+++-}^{2, i j}(s)+T_{+-++}^{2, i j}(s)+T_{++-+}^{2, i j}(s)+T_{-+++}^{2, i j}(s)] \\
& +\frac{\sqrt{3}}{15}[T_{+++0}^{2, i j}(s)+T_{+0++}^{2, i j}(s)+T_{++0+}^{2, i j}(s)+T_{0+++}^{2, i j}(s)\\
&+T_{++-0}^{2, i j}(s)+T_{-0++}^{2, i j}(s)+T_{++0-}^{2, i j}(s)+T_{0-++}^{2, i j}(s)] \\
& +\frac{1}{5}[T_{+00+}^{2, i j}(s)+T_{0++0}^{2, i j}(s)+T_{+00-}^{2, i j}(s)+T_{0-+0}^{2, i j}(s)+T_{+0+0}^{2, i j}(s)\\
&+T_{0+0+}^{2, i j}(s)+T_{0+0-}^{2, i j}(s)+T_{+0-0}^{2, i j}(s)] \\
& +\frac{\sqrt{2}}{5}[T_{+-+0}^{2, i j}(s)+T_{+0+-}^{2, i j}(s)+T_{+-0+}^{2, i j}(s)+T_{0++-}^{2, i j}(s)\\
&+T_{-++0}^{2, i j}(s)+T_{+0-+}^{2, i j}(s)+T_{-+0+}^{2, i j}(s)+T_{0+-+}^{2, i j}(s)] \\
& +\frac{2}{15}[T_{++00}^{2, i j}(s)+T_{00++}^{2, i j}(s)+T_{0000}^{2, i j}(s)]+\frac{2 \sqrt{6}}{15}[T_{+-00}^{2, i j}(s)+T_{00+-}^{2, i j}(s)]\\
&+\frac{2}{5}[T_{+-+-}^{2, i j}(s)+T_{+--+}^{2, i j}(s)]  +\frac{2 \sqrt{3}}{15}[T_{+000}^{2, i j}(s)+T_{00+0}^{2, i j}(s)+T_{0+00}^{2, i j}(s)+T_{000+}^{2, i j}(s)]\ ,
\end{aligned}
\end{equation}
where $i,j=1,2,3$ denote the corresponding outgoing and incoming channels
(corresponding to $J/\psi J/\psi$, $J/\psi\psi(3685)$ and $J/\psi\psi(3770)$, respectively). In practice, it is found that the anomalous enhancement gives a more prominent effect to the $J=2$ amplitude than the $J=0$ amplitude. The amplitude $T^{J,ij}_{L=0}$ is a $K$-
matrix unitarized amplitude obtained from the tree level amplitudes generated from the following contact interaction lagrangian:
\begin{equation}
\begin{aligned}\label{10}
\mathcal{L}_c= & c_1 V_\mu V_\alpha V^\mu V^\alpha+c_2 V_\mu V_\alpha V^\mu V^{\prime \alpha}+c_3 V_\mu V_\alpha^{\prime} V^\mu V^{\prime \alpha}+c_4 V_\mu V^{\prime \mu} V_\alpha V^{\prime \alpha}+c_5 V_\mu V_\alpha V^\mu V^{\prime \prime \alpha} \\
& +c_6 V_\mu V_\alpha^{\prime \prime} V^\mu V^{\prime \prime \alpha}+c_7 V_\mu V^{\prime \prime \mu} V_\alpha V^{\prime \prime \alpha}+c_8 V_\mu V_\alpha^{\prime} V^\mu V^{\prime \prime \alpha}+c_9 V_\mu V^{\prime \mu} V_\alpha V^{\prime \prime \alpha},
\end{aligned}
\end{equation}
The production amplitudes $F^J_i$ are parameterized as,
\begin{equation}
    F^J_i(s) = \sum^3_{k = 1} \alpha_k(s) T_L^{J,ki}(s)~,
\end{equation}
where $\alpha_k(s)$ are real polynomial functions in general and are set to be constants here, and further we set  $\alpha_1(s)^2 = 1$.  Further, 
\begin{equation}
    \frac{d Events_i}{d\sqrt{s}} = N_i~p_{i}(s)~|F_i|^2~,
\end{equation}
where $p_{i}(s)$ refers to the abs of three momentum for corresponding channel. According to partial wave convention, for $J=0$ and $J=2$ case, they have a total scale factor that~\cite{Liang:2021fzr, Zhou:2022xpd}
\begin{equation}
    |F_i|^2 = |F_i^{J=0}|^2 + 5~|F_i^{J=2}|^2\ .
\end{equation}
The fit is  overdone since there are many parameters. One solution is  shown in Fig.~\ref{fit}, and the fit parameters are listed in Tab.~\ref{tab1} for illustration.  During the fit many solutions are examined, nevertheless  it is found that the triangle diagram contributions are all small. It is not totally clear to us why it  behaves like such but one possible reason could be that the peak position through anomalous threshold contribution, as shown in Fig.~\ref{fig:3}, is roughly 60 -- 80MeV above the $X(6900)$ peak, and hence the fit does not like it. One possible way to rescue the problem may be to adopt another parameterization in which the background contributions are more flexible to be tuned, hence the interference between the background and the anomalous enhancement can lead to the shift of the peak position by a few tens of MeV.   We leave this investigation for future studies.

\begin{table}[h]
\centering
\begin{tabular}{lcccccc}
   \toprule
    Parameter& $\chi^2/d.o.f$ & $g'$ & $N_1$ & $N_2$ & $\alpha_2$ & $\alpha_3$ \\
   \midrule
    Fit&  $0.93$  & $-18.3\pm 1.6$& $23\pm 10$ & $0.34\pm 0.23$ & $-4.94\pm 0.02$ & $3.97\pm 0.05$ \\
   \bottomrule
\end{tabular}

\begin{tabular}{ccccc}
   \toprule
    $c_5$& $c_6$& $c_7$& $c_8$& $c_9$\\
   \midrule
    $11.34\pm 0.05$ & $50.79\pm0.05$ & $-35.01\pm0.19$&$-64.71\pm 0.07$ &$1.529\pm 0.002$\\
   \bottomrule
\end{tabular}
\caption{The fit parameters of Fig.~\ref{fit}. The $c_i$ parameters are defined in Eq.~(\ref{10}). In this fit,
we set $c_1,\cdots , c_4$ to be negligible.}
\label{tab1}
\end{table}

\section*{Acknowledgements}
We would like to thank De-Liang Yao and Ling-Yun Dai for very helpful discussions.
This work is supported in part by National Nature Science Foundations
of China under Contract Number 12335002.


\begin{thebibliography}{}

 
    \bibitem{6900}
       R. Aaij \textit{et al.} (LHCb Collaboration), Sci.Bull. \textbf{65} (2020) 23, 1983.
       
   \bibitem{CMS} CMS collaboration, CMS-PAS-BPH-21-003 (2022).

    \bibitem{ATLAS:2022hhx} ATLAS collaboration, ATLAS-CONF-2022-040 (2022).
    
    \bibitem{Lu:2023ccs}
    Y. Lu {\it et al.},   Phys. Rev. D{\bf 107} (2023) 094006.

    \bibitem{Cao:2020gul} 
    Q. F. Cao $et$ $al.$, 
     Chin. Phys. \textbf{C} 45 (2021) 10, 103102.


   \bibitem{pole} D.~Morgan, Nucl. Phys. A \textbf{543} (1992) 632.

 
   \bibitem{Zhang:2009bv} O.~Zhang, C.~Meng and H.~Q.~Zheng,
  Phys. Lett. B \textbf{680} (2009) 453.

 
  \bibitem{Dai:2012pb} L.~Y.~Dai, M.~Shi, G.~Y.~Tang and H.~Q.~Zheng,
   Phys. Rev. D \textbf{92} (2015) 014020.

   \bibitem{X3900} Q.~R.~Gong, Z.~H.~Guo, C.~Meng, G.~Y.~Tang, Y.~F.~Wang and H.~Q.~Zheng,
   Phys. Rev. D \textbf{94} (2016) 114019.

   \bibitem{Cao:2019wwt} Q.~F.~Cao, H.~R.~Qi, Y.~F.~Wang and H.~Q.~Zheng,
   Phys. Rev. D \textbf{100} (2019) 054040.
   
   \bibitem{Liang:2021fzr} Z. R. Liang, X. Y. Wu, D. L. Yao,  Phys. Rev. D\textbf{104} (2021)  034034. 
  
   \bibitem{Zhou:2022xpd} Q. Zhou, D. Guo, S. Q. Kuang, Q. H. Yang and L. Y. Dai, Phys. Rev. D \textbf{106} (2022) 11.

   \bibitem{Mandelstam} S. Mandelstam, Phys. Rev. Lett. 4 (1960) 84.

\bibitem{Braaten} E. Braaten, H. W. Hammer,  Phys. Rep.  \textbf{428} (2006) 259.
\bibitem{Hammer} H. W. Hammer, C. Ji and D. R. Phillips, J. Phys.  G\textbf{44} (2017) 103002. 

\bibitem{PDG}  M. Tanabashi {\it et al.} (Particle Data Group), Phys. Rev. D \textbf{98} (2018) 030001.


\bibitem{Coito:2017ppc} S. Coito and F. Giacosa, Nucl. Phys. A \textbf{981} (2019) 38. 


\end{thebibliography}
\end{document}